\documentclass[12pt,titlepage]{article}
\usepackage{epsfig}
\usepackage{lscape}
\usepackage{amsmath,epsf,color,colordvi,shadow,pifont}

\usepackage{amsmath}
\usepackage{url}
\usepackage{graphicx}
\usepackage{amssymb}
\usepackage{amsmath,amsfonts,amssymb} 
\usepackage{graphicx} 
\usepackage{lmodern}
\usepackage{hyperref} 
\newfont{\futharknormal}{futhol10} 

\begin{document}

\begin{titlepage}
\begin{center}
{\LARGE {\bf  Dynamics of the four kinds of \\ Trapping Horizons \\ \& \\ Existence of Hawking Radiation}}
 \\
\vskip 1cm

{\large Alexis Helou\footnote{alexis.helou@apc.univ-paris7.fr}} \\
{\em AstroParticule et Cosmologie, 
Universit\'e Paris Diderot, CNRS, CEA, Observatoire de Paris, Sorbonne Paris 
Cit\'e} \\
{B\^atiment Condorcet, 10, rue Alice Domon et L\'eonie Duquet,\\
F-75205 Paris Cedex 13, France}
\end{center}

\centerline{ {\bf Abstract}} We work with the notion of apparent/trapping horizons for spherically symmetric, dynamical spacetimes: these are quasi-locally defined, simply based on the behaviour of congruence of light rays. We show that the sign of the dynamical Hayward-Kodama surface gravity is dictated by the inner/outer nature of the horizon. Using the tunneling method to compute Hawking Radiation, this surface gravity is then linked to a notion of temperature, up to a sign that is dictated by the future/past nature of the horizon. Therefore two sign effects are conspiring to give a positive temperature for the black hole case and the expanding cosmology, whereas the same quantity is negative for white holes and contracting cosmologies. This is consistent with the fact that, in the latter cases, the horizon does not act as a separating membrane, and Hawking emission should not occur.

\indent 

\vfill  
\end{titlepage}

\tableofcontents

\section{Introduction}

In the 1960's, fifty years after the birth of General Relativity, the field of gravitation theory underwent a period of great renewal. New ideas by Penrose and many others led to theorems on spacetime singularities (see the recent review \cite{Senovilla:2014gza}) and to the elaboration of new tools to describe the causality of spacetimes: the so-called Penrose-Carter diagrams. These spacetime diagrams were first used to represent simple, ideally static situations, such as the Schwarzschild geometry for black holes and the de Sitter one for cosmology. In these situations where there is no change in time and where future infinity can be absolutely predicted, a very beautiful and powerful notion arises: the concept of event horizon. It is a line in the Penrose diagram, which separates all the points (the events) into two categories, depending on whether the observer will or will not be able to receive signals from them. The event horizon is thus an absolute horizon by definition: no information can cross it, at least classically. However, Hawking showed in the seventies that a black hole event horizon could emit radiation semi-classically, and bestowed a thermal spectrum with positive temperature $T$ on the horizon \cite{Hawking:1974sw}. The temperature is linked to the surface gravity of the black hole, $T=\kappa/2\pi$. This result was soon generalized to de Sitter Universe and the cosmological event horizon \cite{Gibbons:1977mu}. Since this seminal work, a lot of interest has been driven towards cosmological horizons and their thermodynamics, but there have always been some confusion about the sign of the temperature. Indeed, the geometrical surface gravity is negative for de Sitter, and one may wonder whether one should use an absolute value in the definition of the temperature, $T=\lvert \kappa\rvert/2\pi$, or choose a positive definition for the surface gravity $\kappa$. 

The present paper addresses this issue, but in the more general context of spherically symmetric, \emph{dynamical} spacetimes. We go beyond the static Schwarzschild and de Sitter cases, and consider general metrics for dynamical black holes and dynamically expanding cosmologies. This is motivated by the fact that physically realistic systems would not be static: black holes accrete matter and evaporate, and the Friedmann-Lemaître-Robertson-Walker (FLRW) cosmology has proven to be a very accurate dynamical description of our Universe. Moreover our formalism, taken from \cite{Hayward:1993wb}, can describe white holes and contracting cosmologies as well. In this general context, we aim at explaining how the Hayward-Kodama surface gravity may be linked to a notion of temperature. 
We find a positive temperature for both the black hole and expanding cosmology cases, without any need to put absolute values by hand. On the other hand however, we find a negative temperature for white holes and contracting cosmologies. We give a physical argument to explain this fact, using the picture of Hawking radiation coming from pairs of particle-antiparticle created at the horizon.  For a black hole, the horizon does act as a separating surface for the two members of the pair, as well as for an expanding cosmology. However, in the other two situations, we realize that the horizon has no separating effect on pairs of particles. In these cases, we conclude that Hawking radiation should not occur.

\section{Foreword}

 Note that, in order to go from a static description to a dynamical one, we need to let go of one of the most prominent tools in the work of Penrose, the event horizon. Indeed, in a dynamical framework there is no way of predicting which events will remain hidden forever to the observer. We need another definition for the horizon, which would not be global or teleological. Somehow paradoxically, this is provided by another tool due to Penrose, \emph{i.e.} trapped surfaces. These define the notion of apparent or trapping horizons, using the quasi-local contraction/expansion of bundles of light-rays. 
 
We will use the terms apparent/trapping horizons interchangeably \footnote{The two notions are often taken as synonymous in the literature \cite{Poisson}. Strictly speaking, the trapping horizon is a 3D tube obtained by stacking together 2D apparent horizons \cite{Hawking:1973uf}.}. There can be several types of trapping horizons, named within the nomenclature of Hayward \cite{Hayward:1993wb}: future/past, inner/outer \footnote{We will however only consider situations where the notions of ``inwards'' and ``outwards'' can be physically distinguished, so that we will not use Hayward's convention that $\theta_+$ is always the expansion that vanishes on the horizon. This is true for the black hole case, but for expanding cosmologies, it is $\theta_-$ that vanishes on the apparent horizon. Except for that convention, we follow \cite{Hayward:1993wb}.}.
 
 The precise aim of this article is to show that the future/past, inner/outer character of the horizon has a two-fold effect on the sign of the temperature of the horizon. In Section \ref{section_past_horizons} and Section \ref{section_future_horizons}, we establish the link between the sign of Hayward-Kodama surface gravity and the inner/outer nature of the horizon (which is given by the sign of a certain Lie derivative). We obtain that all outer horizons have positive surface gravity, whereas inner horizons have negative $\kappa$. The discussion is split into two sections, for past horizons and future ones, but this is only for computational reasons, since each case comes with its own metric. This separation has no deep meaning: the conclusion is the same whatever the future/past character. This is the first step, the first sign effect. 
 The second sign effect comes from the computation of the temperature of Hawking radiation in the tunneling method of Parikh and Wilczek \cite{Parikh:1999mf}. In Section \ref{section_Hawking_rad_from_tunneling}, we indeed show that future/past horizons, which are described by two metrics only differing by a sign, also differ by the same sign in the relation between the temperature of the thermal spectrum and the surface gravity, $T\propto \pm \kappa$ respectively.
 
 When both sign effects are taken into account, we find that only future-outer and past-inner horizons have positive temperature, whereas $T$ is negative for past-outer and future-inner configurations. In Section \ref{section_physical_arguments}, we give examples for each type of trapping horizons, and explain why the horizon can act as a separation for particles in some configurations, and cannot in some others. Section \ref{section_conclusion} summarizes our conclusions.

\section{Past Horizons: Retarded Eddington-Finkelstein metric}
 \label{section_past_horizons}
\subsection*{Spherical symmetry and dynamical surface gravity}
Any spherically symmetric geometry can locally be written in dual-null coordinates $(x_+, x_-)$:
\begin{equation}
ds^2 = -e^{2\Psi}C dx_+dx_- +  R^2d\Omega^2 \ .
 \label{eq_duall_null_metric}
\end{equation}
Here, $C$ and $\Psi$ are two functions of $x_+$ and $x_-$, and $R$ is the areal radius (radius of the 2-spheres of symmetry). Another possibility is to write this general metric in a coordinate system where the Kodama time {\futharknormal X} \footnote{{\futharknormal X} is a rune of the Scandinavian futhark alphabet. It is named Thurisaz, or Thurs.} is explicit \cite{Kodama:1979vn},\cite{Abreu:2010ru}:
\begin{equation}
ds^2 = -e^{2\Psi}Cd\text{\futharknormal X}^2 + \frac{1}{C}dR^2 + R^2d\Omega^2 \ .
 \label{eq_Kodama_time_metric}
\end{equation}
We have here used the following transformations:
\begin{equation}
\begin{cases}
 dx_- =  d\text{\futharknormal X}  - (e^{\Psi}C)^{-1}dR \ ,  \\ \nonumber
 dx_+ =  d\text{\futharknormal X}  + (e^{\Psi}C)^{-1}dR \ . \nonumber
\end{cases}
\end{equation}
The signature of this metric is $(-,+,+,+)$ wherever $C$ is positive, which we will take as the indicator that we are in a normal region of spacetime (Minkowski-like). On the other hand, $C$ being negative will be the marker of a trapped region (future or past). The transition between two regions, where $C$ vanishes, is by definition the apparent/trapping horizon. 

We notice that the metric Eq.\eqref{eq_Kodama_time_metric} is therefore singular on the apparent horizon. In order to get rid of this coordinate singularity, one has the choice of going either to the advanced Eddington-Finkelstein coordinates $x_+=\eta +\chi ,$ or to the retarded EF coordinates $x_-=\eta - \chi$ (where $\eta$ and $\chi$ are conformal time and distance). The advanced coordinates are fit to describe a future horizon (\emph{e.g.} a black hole horizon), while the retarded coordinates are used for past horizons (\emph{e.g.} a cosmological, Big Bang horizon). Let us choose either one of them, for example the retarded Eddington-Finkelstein coordinates:
\begin{equation}
ds^2 = -e^{2\Psi}Cdx_-^2 - 2e^{\Psi}dx_-dR + R^2d\Omega^2 \ .
 \label{eq_retarded_EF_metric}
\end{equation}
The metric and inverse metric tensors read:
\begin{equation}
g_{\mu\nu}=
\begin{pmatrix} 
  -e^{2\Psi}C     & -e^{\Psi}\\ 
  -e^{\Psi}       & 0 
\end{pmatrix}   \ ,
\end{equation}
\begin{equation}
g^{\mu\nu}=
\begin{pmatrix} 
  0             & -e^{-\Psi}\\ 
  -e^{-\Psi}    & C
\end{pmatrix} \ .
\end{equation}
This coordinate chart covers both sides of a past apparent/trapping horizon, where the parameter $C$ vanishes. There is no coordinate singularity at the horizon (the principle of EF coordinates). This can describe a Friedmann-Lemaître-Robertson-Walker spacetime, identifying the parameters as follows:
\begin{equation}
e^{\Psi}=\frac{a}{\sqrt{1-kr^2}+RH} \ ,
\end{equation}
\begin{equation}
e^{2\Psi}C=a^2\frac{\sqrt{1-kr^2}-RH}{\sqrt{1-kr^2}+RH} \ , 
\end{equation}
\begin{equation}
C=1-kr^2-R^2H^2 \ ,
 \label{eq_def_C_cosmo}
\end{equation}
where $a$ is the scale-factor, $k$ the curvature of space, and $H$ the Hubble parameter. The physical, areal radius $R$ can be written as $R=a r$. $C$ vanishes at the apparent horizon $R_A = (H^2 + k/a^2)^{-1/2}$, which boils down to the Hubble Sphere for flat space ($k=0$). However we do not need to specialize to FLRW here, so we go back to the general case of a past horizon, within Eq.\eqref{eq_retarded_EF_metric}.

Following the Hayward-Kodama \cite{Hayward:1997jp} prescription for defining a surface gravity in dynamical spacetimes (\emph{i.e.} deprived of a time-translational Killing field), we write the surface gravity as:
\begin{align}
\kappa &= \frac{1}{2\sqrt{-\gamma}}\partial_i[\sqrt{-\gamma}\gamma^{ij}\nabla_j R] \\ \nonumber
&=\frac{1}{2}\left( \partial_R C + C\partial_R \Psi \right) \ ,
 \label{eq_surface_grav_EF_outgoing}
\end{align}
where $\gamma_{i j}$ is the metric of the 2-space orthogonal to the spheres of symmetry:
\begin{equation}
ds^2 = \gamma_{i j}(x)dx^i dx^j + R^2(x)d\Omega^2 \ .
 \label{eq_metric_spherical}
\end{equation}
Now, evaluating $\kappa$ on the apparent/trapping horizon (where C vanishes):
\begin{equation}
\kappa_{\mathcal{H}} = \left . \frac{ \partial_R C}{2}  \right \vert_{\mathcal{H}} \ .
\end{equation}
We have shown in \cite{Binetruy:2014ela} that this surface gravity is negative for a FLRW Universe with physical matter content (matter, radiation, and even dark energy). The limit case is obtained for radiation, where $\kappa_{\mathcal{H}} =0$. Beyond that (for state parameter $w>1/3$), the apparent horizon is no longer of the past-inner type, but is past-outer, as explained in \cite{Helou:2015yqa}. We will therefore restrict our study of expanding cosmologies to past-inner horizons \footnote{However, a stiff-matter Universe with equation of state $p=\rho$ is an example of special expanding cosmology with past-outer horizon.}.

\subsection*{Null normal vectors}
In order to compute the expansion of null geodesic congruence, we will be looking for the vector field $l^a$, null and orthogonal to the 2-spheres of symmetry, which enters in the geodesic equation:
\begin{equation}
\frac{dl^a}{d \lambda} + \Gamma^{a}_{\phantom{a} b c}l^bl^c = 0 \ .
\end{equation}
It is defined by:
\begin{equation}
l^a = \frac{dx^a}{d\lambda} \ ,
\end{equation} 
where $\lambda$ is an affine parameter, chosen as $\lambda = x_-$. Then $l^0=1$, and for a radially moving photon:
\begin{equation}
ds^2 = 0 = -e^{2\Psi}C dx_-^2 - 2 e^{\Psi}dx_-dR \ .
\end{equation}
Therefore the second component is:
\begin{equation}
l^1 = \frac{dR}{dx_-} = - \frac{e^{\Psi}C}{2} \ ,
 \label{eq_l1_retarded}
\end{equation}
for $dx_- \neq 0$.
So finally:
\begin{align}
l^a &= (1 , - \frac{e^{\Psi}C}{2}, 0, 0) \ , \nonumber \\
l_a &= (-\frac{e^{2\Psi}C}{2}, -e^{\Psi}, 0, 0) \ ,
\end{align}
and we check that $l^al_a = 0 \ .$

Note here that we only have one available future direction for the light rays, for $dx_- \neq 0$ (\emph{i.e.} $x_+=cst$). This is because the metric is non-diagonal. In the normal regions where C is positive, it corresponds to the ingoing direction, since $\frac{dR}{dx_-}\leq 0$, see Eq.\eqref{eq_l1_retarded}. Thus $l^a$ is the future-ingoing, normal null vector. The other future direction will be given by $dx_-=0$.

Therefore the future-outgoing, orthogonal null vector $k^a$ is:
\begin{equation}
k^a = (0, k^1, 0, 0) \ .
\end{equation}
With the normalisation $l_ak^a = -2$:
\begin{equation}
 - e^{\Psi} k^1 = -2 \Rightarrow k^1 =  2e^{-\Psi} \ .
\end{equation}
So the second null vector reads:
\begin{align}
k^a &= (0, 2e^{-\Psi}, 0, 0) \ , \nonumber \\
k_a &= (-2, 0, 0, 0) \ .
\end{align}
We indeed check that $k^ak_a=0$.

\subsection*{Expansion of null geodesic congruences}
From the null vectors above we can define the induced metric on the 2-spheres:
\begin{equation}
h_{ab} = g_{ab} + \frac{1}{2}(k_al_b + k_bl_a)
= \begin{pmatrix} 
  0 & 0 & 0 & 0\\ 
  0 & 0 & 0 & 0\\
  0 & 0 & R^2 & 0\\
  0 & 0 & 0 & R^2sin^2(\theta)
\end{pmatrix} \ .
\end{equation}
The expansion of the ingoing null congruences is by definition:
\begin{align}
\theta_- &= h^{cd}\nabla_c l_d \nonumber \\
&= h^{cd}(\partial_c l_d - \Gamma^{a}_{\phantom{a} c d} l_a) \nonumber \\
& = -h^{22}\Gamma^{0}_{\phantom{0} 2 2} l_0 -h^{22}\Gamma^{1}_{\phantom{1} 2 2} l_1 -h^{33}\Gamma^{0}_{\phantom{0} 3 3} l_0  -h^{33}\Gamma^{1}_{\phantom{1} 3 3} l_1 \ ,
\end{align}
which yields:
\begin{equation}
\theta_- = - \frac{e^{\Psi}C}{R}  \ .
 \label{eq_theta-retarded}
\end{equation}
At the apparent horizon, we know that C vanishes, thus we indeed have that $\theta_-$ vanishes on the apparent horizon (by definition of a past horizon). The important thing to notice here is that the expansion changes sign at horizon crossing.  Let us compute the expansion on the other future direction, and check that it has a constant sign at horizon crossing. The expansion of the outgoing null congruences reads:
\begin{align}
\theta_+ &= h^{cd}\nabla_c k_d \nonumber \\
&= h^{cd}(\partial_c k_d - \Gamma^{a}_{\phantom{a} c d} k_{a}) \nonumber \\
& = -h^{22}\Gamma^{0}_{\phantom{0} 2 2} k_0 -h^{22}\Gamma^{1}_{\phantom{1} 2 2} k_1 -h^{33}\Gamma^{0}_{\phantom{0} 3 3} k_0  -h^{33}\Gamma^{1}_{\phantom{1} 3 3} k_1 \ .
\end{align}
We obtain:
\begin{equation}
\theta_+ =  \frac{4e^{-\Psi}}{R}  \ ,
 \label{eq_theta+retarded}
\end{equation}
which indeed is always positive. Using Eqs.\eqref{eq_theta-retarded} and \eqref{eq_theta+retarded}, one may draw the Bousso wedges (light-cone directions where $\theta<0$) on the Penrose diagram of this spacetime. We indeed recover the usual diagram for FLRW (Figure \ref{fig_FLRW}).

\subsection*{Lie derivative of the expansion: inner and outer past horizons}
With the above setup in hands, we can describe both past-inner and past-outer horizons. The only prescription so far is that $\theta_-$ should vanish on the horizon, which is a by-product of the choice of the retarded EF coordinates. Now we have two possibilities for the change of sign of $\theta_-$ at the horizon. Either:
\begin{equation}
\theta_-=0, \quad  \theta_+>0  \quad \text{ and } \quad \mathcal{L}_+ \theta_- > 0 \ ,
\end{equation}
which is the definition of a past-inner trapping horizon \cite{Hayward:1993wb}. Or:
\begin{equation}
\theta_-=0, \quad  \theta_+>0  \quad \text{ and } \quad \mathcal{L}_+ \theta_- < 0 \ ,
\end{equation}
which is the definition of a past-outer trapping horizon.
$\mathcal{L}_+ \theta_- $ denotes the Lie derivative of the expansion along the null direction $k^a$ where $x_+$ varies, \emph{i.e.} where $x_-=cst$. For a given type of horizon (say past-inner), this Lie derivative has constant sign. Since the null vector $k^a$ writes, in the different coordinate systems, as:
\begin{equation}
k^a = \left(0, \left . \frac{\partial R}{ \partial x_+ } \right\vert_{x_-}=2e^{-\Psi}, 0, 0\right)_{(x_-,R,\theta,\phi)}= (1, 0, 0, 0)_{(x_+,x_-,\theta,\phi)} \ ,
 \label{eq_retarded_null_vector}
\end{equation}
we may compute the Lie derivative.
 \begin{align}
\mathcal{L}_+ \theta_- &= k^a\partial_a \theta_- = \left . \frac{\partial}{ \partial x_+ } \left(- \frac{e^{\Psi}C}{R} \right) \right\vert_{x_-} \\ \nonumber
 &=  \left . \frac{\partial }{ \partial R } \left(- \frac{e^{\Psi}C}{R} \right) \right\vert_{x_-} \times \left . \frac{\partial R}{ \partial x_+ } \right\vert_{x_-}  \\ \nonumber
 &= -\frac{1}{R^2} \left . \left( e^{\Psi}CR \partial_R \Psi + e^{\Psi}R \partial_R C - e^{\Psi}C  \right)\right\vert_{x_-} \times \left . \frac{\partial R}{ \partial x_+ } \right\vert_{x_-}  \ .
 \label{eq_lie_deriv}
\end{align}
We can expand this expression around the horizon, where $C=0$:
\begin{equation}
\left . \mathcal{L}_+ \theta_- \right\vert_{\mathcal{H}} = - \frac{e^{\Psi}}{R} \left . \frac{\partial R}{ \partial x_+ } \right\vert_{x_-} \left . \partial_R C  \right\vert_{x_-}  + \mathcal{O}(C) \ .
 \label{eq_lie_deriv_expand}
\end{equation} 
Since $\kappa_{\mathcal{H}} =  \frac{1}{2}  \partial_R C \vert_{x_-}$:
\begin{equation} 
\left . \mathcal{L}_+ \theta_- \right\vert_{\mathcal{H}} \sim - \frac{2e^{\Psi}}{R} \left . \frac{\partial R}{ \partial x_+ } \right\vert_{x_-} \times \kappa_{\mathcal{H}} \ .
 \label{eq_lie_deriv_expand2}
\end{equation}
Let us focus on the sign of $\left . \frac{\partial R}{ \partial x_+ } \right\vert_{x_-}$. This is the variation of the radius of the 2-spheres of symmetry along a future-outgoing null ray. In the present case of past horizons, $\theta_+$ is always positive, $R$ is always increasing in the future-outgoing null direction. The derivative in question is therefore always positive, which checks Eq.\eqref{eq_retarded_null_vector}. We finally obtain that $\mathcal{L}_+ \theta_- $ has a sign opposite to $\kappa_{\mathcal{H}}$, which is therefore also constant for a given type of past horizon. A past-inner horizon will have negative surface gravity, and a past-outer one will have a positive $\kappa_{\mathcal{H}}$.

\vspace{2mm}
However, the Lie derivative of the same expansion but in the other null direction $l^a$, $\mathcal{L}_- \theta_- \ ,$ changes sign depending on the timelike/ lightlike/ spacelike nature of the horizon:
\begin{equation}
\mathcal{L}_- \theta_- = l^a\partial_a \theta_- = \left . \frac{\partial}{ \partial x_- } \left(- \frac{e^{\Psi}C}{R} \right) \right\vert_{x_+}  \ .
 \label{eq_lie_deriv--}
\end{equation}
Near the horizon, an expansion of the above reads:
\begin{equation}
\left . \mathcal{L}_- \theta_- \right\vert_{\mathcal{H}} = - \frac{e^{\Psi}}{R} \left . \frac{\partial C}{\partial x_-}  \right\vert_{x_+} + \mathcal{O}(C) \ .
 \label{eq_lie_deriv--_expand}
\end{equation}
This Lie derivative enters the definition of $\alpha$ \cite{Dreyer:2002mx}:
\begin{equation}
\alpha = \frac{ \mathcal{L}_- \theta_- }{ \mathcal{L}_+ \theta_- }   \ ,
 \label{eq_lie_deriv--_expand2}
\end{equation}
which sign gives the causal nature of the horizon, timelike/null/spacelike for negative/null/positive $\alpha$ respectively. All of the above section is applicable for both inner and outer past horizons. Let us now move to future horizons.

\section{Future Horizons: Advanced Eddington-Finkelstein metric}
 \label{section_future_horizons}
The computations of the previous section can easily be transposed to advanced Eddington-Finkelstein metric:
\begin{equation}
ds^2 = -e^{2\Psi}C dx_+^2 + 2e^{\Psi}dx_+dR + R^2d\Omega^2 \ .
 \label{eq_EF_ingoing_metric}
\end{equation}
The surface gravity has exactly the same form as for past horizons:
\begin{equation}
\kappa_{\mathcal{H}} = \frac{1}{2} \left . \partial_R C\right \vert_{x_+} \ .
 \label{eq_surf_grav_future}
\end{equation}
However now it is the outgoing expansion that vanishes at the horizon:
\begin{equation}
\theta_+ = + \frac{e^{\Psi}C}{R} \ .
 \label{eq_theta-advanced}
\end{equation}
For future horizons, we also have two possibilities:
\begin{equation}
\theta_+=0, \quad  \theta_-<0  \quad \text{ and } \quad \mathcal{L}_- \theta_+ > 0 \ ,
\end{equation}
which is the definition of a future-inner trapping horizon. And:
\begin{equation}
\theta_+=0, \quad  \theta_-<0  \quad \text{ and } \quad \mathcal{L}_- \theta_+ < 0 \ ,
\end{equation}
which is the definition of a future-outer trapping horizon.
Let us now compute the Lie derivative of the expansion $\theta_+$ in the null direction $l^a$ where $x_-$ varies:
 \begin{align}
\mathcal{L}_- \theta_+ &= l^a\partial_a \theta_+ =\left . \frac{\partial}{ \partial x_- } \left( \frac{e^{\Psi}C}{R} \right) \right\vert_{x_+} \nonumber \\
 &= \left . \frac{\partial}{ \partial R } \left( \frac{e^{\Psi}C}{R} \right) \right\vert_{x_+} \times \left . \frac{\partial R }{ \partial x_- }  \right\vert_{x_+} \nonumber \\
 &= \frac{1}{R^2} \left . \left( e^{\Psi}CR \partial_R \Psi + e^{\Psi}R \partial_R C - e^{\Psi}C  \right)\right\vert_{x_+} \times \left . \frac{\partial R }{ \partial x_- }  \right\vert_{x_+}  \ .
 \label{eq_lie_deriv_future}
\end{align}
Around the horizon, where $C$ vanishes, this can be expanded as:
\begin{equation}
\left . \mathcal{L}_- \theta_+ \right\vert_{\mathcal{H}} =  \frac{e^{\Psi}}{R} \left . \partial_R C  \right\vert_{x_+} \times \left . \frac{\partial R }{ \partial x_- }  \right\vert_{x_+}  + \mathcal{O}(C) \ ,
 \label{eq_lie_deriv_expand_future}
\end{equation}
and, identifying the surface gravity in Eq.\eqref{eq_surf_grav_future}:
\begin{equation}
\left . \mathcal{L}_- \theta_+ \right\vert_{\mathcal{H}} \sim  \frac{2e^{\Psi}}{R} \left . \frac{\partial R }{ \partial x_- }  \right\vert_{x_+} \times \kappa_{\mathcal{H}} \ .
 \label{eq_lie_deriv_expand2_future}
\end{equation}
We need to study the sign of $\left . \frac{\partial R }{ \partial x_- }  \right\vert_{x_+}$, which is the derivative of the areal radius along a future-ingoing light ray. Since for future horizons, the expansion $\theta_-$ is always negative, this derivative is also negative. Thus $\mathcal{L}_- \theta_+ $ has a sign opposite to that of $\kappa_{\mathcal{H}}$, just as for past horizons. A future-outer horizon has positive surface gravity, while it is negative for a future-inner horizon.

\vspace{2mm}
Now let us look at the other Lie derivative, $\mathcal{L}_+ \theta_+$.
\begin{equation}
\mathcal{L}_+ \theta_+ = \left . \frac{\partial}{ \partial x_+ } \left( \frac{e^{\Psi}C}{R} \right) \right\vert_{x_-}  \ .
 \label{eq_lie_deriv++}
\end{equation}
Expanding near the horizon:
\begin{equation}
\left . \mathcal{L}_+ \theta_+ \right\vert_{\mathcal{H}} =  \frac{e^{\Psi}}{R} \left . \frac{\partial C}{\partial x_+}  \right\vert_{x_-} + \mathcal{O}(C) \ .
 \label{eq_lie_deriv++_expand}
\end{equation}
Here again we may define the ratio $\alpha$:
\begin{equation}
\alpha = \frac{ \mathcal{L}_+ \theta_+ }{ \mathcal{L}_- \theta_+ }   \ ,
 \label{eq_lie_deriv++_expand2}
\end{equation}
which is  negative/null/positive for timelike/null/spacelike horizons. Note that in this section, we have not specified the inner/outer character of the future horizon. Everything is valid for both cases.

In Section \ref{section_past_horizons} and Section \ref{section_future_horizons}, we have shown that the surface gravity of the horizon has a sign opposite to that of the Lie derivative of the vanishing expansion. This is true regardless of the future/past character of the horizon, since both sections recover the same conclusion. Since this Lie derivative enters the very definition of inner/outer horizons, we can conclude that:
\begin{itemize}
 \item  an inner horizon has negative surface gravity,
 \item  an outer horizon has positive surface gravity.
\end{itemize}
This is the first sign effect that we establish. In the next section, we will show that a second sign effect arises from the future/past nature of the horizon, in the relation $T=\pm \kappa_{\mathcal{H}}$.

\section{Hawking Radiation from Tunneling}
 \label{section_Hawking_rad_from_tunneling}
 
In this section, we recall the results for the computation of Hawking radiation using the tunneling method of Parikh and Wilczek \cite{Parikh:1999mf}. The original computations may be found in \cite{Hayward:2008we} for the black hole case, and in \cite{Binetruy:2014ela} and \cite{Helou:2015yqa} for the case of an expanding cosmology.

\bigskip
\noindent
\begin{minipage}[t]{.5\textwidth}
\raggedright
 \begin{center}
 \begin{itemize}
 \item Future Horizon \\ $\rightarrow$ Advanced Eddington-Finkelstein:
 \begin{eqnarray}
 ds^2 = -e^{2\Psi}Cdx_+^2 + 2e^{\Psi}dx_+dR  \nonumber
 \\ + R^2d\Omega^2 \ . \nonumber
 \end{eqnarray}
 \end{itemize}
  ($x_+=\eta+\chi$)
 \end{center} 
\end{minipage}
\hfill
\noindent
\begin{minipage}[t]{.5\textwidth}
\raggedleft
 \begin{center}
 \begin{itemize}
 \item Past Horizon \\ $\rightarrow$ Retarded \\  Eddington-Finkelstein:
 \begin{eqnarray}
 ds^2 = -e^{2\Psi}Cdx_-^2 - 2e^{\Psi}dx_-dR  \nonumber
 \\ + R^2d\Omega^2 \ . \nonumber
 \end{eqnarray}
 \end{itemize}
  ($x_-=\eta-\chi$)
 \end{center} 
\end{minipage}
\begin{center}
\item -Kodama vector: $K^a=(e^{-\Psi} ; 0 ; 0 ; 0) \ .$
\item -Surface gravity: $\kappa_{\cal H} = \frac{1}{2} \left. \partial_R C\right|_{\cal H} \ .$
\item -BKW approximation of tunneling probability for a massless scalar field $\phi=\phi_0 \exp(i\mathcal{I})$:
\begin{equation}
\Gamma \propto exp\left(-2\frac{Im\mathcal{I}}{\hbar}\right) \ .
\end{equation}
\item -Equation of Motion is Hamilton-Jacobi Equation:
\begin{equation}
g^{ab} \nabla_a \mathcal{I} \nabla_b \mathcal{I}= 0 \ .
\end{equation}
\end{center}

\bigskip
\noindent
\begin{minipage}[t]{.5\textwidth}
\raggedright
 \begin{center}
 \begin{itemize}
 \item Action: \\$\mathcal{I}=\int\partial_{x_+}(\mathcal{I})dx_++ \int\partial_{R}(\mathcal{I})dR \ .$
 \item Kodama energy: \\$\omega=-K^a\nabla_a \mathcal{I} = -e^{-\Psi}\partial_{x_+}(\mathcal{I}) \ .$ 
 \item Wave number: $k=\partial_{R} \mathcal{I} \ .$
 \item EoM: $k(Ck-2\omega) = 0 \ .$
 \\$\rightarrow$ $k=0$ (ingoing solution).
 \\$\rightarrow$ $k=+2\omega/C$ (outgoing).
 \item Outgoing solution has a pole and contributes to: 
 \begin{equation}
  Im\mathcal{I} = +\frac{\pi\omega}{\kappa_{\cal H}}\ . \nonumber
 \end{equation}
 \end{itemize}
 \end{center} 
\end{minipage}
\hfill
\noindent
\begin{minipage}[t]{.5\textwidth}
\raggedleft
 \begin{center}
 \begin{itemize}
 \item Action: \\$\mathcal{I}=\int\partial_{x_-}(\mathcal{I})dx_- + \int\partial_{R}(\mathcal{I})dR\ .$
 \item Kodama energy: \\$\omega=-K^a\nabla_a \mathcal{I} = -e^{-\Psi}\partial_{x_-}(\mathcal{I}) \ .$
 \item Wave number: $k=\partial_{R} \mathcal{I} \ .$
 \item EoM: $k(Ck+2\omega) = 0 \ .$
 \\$\rightarrow$ $k=0$ (outgoing solution).
 \\$\rightarrow$ $k=-2\omega/C$ (ingoing).
 \item Ingoing solution has a pole and contributes to: 
 \begin{equation}
  Im\mathcal{I} =  -\frac{\pi\omega}{\kappa_{\cal H}}\ .\nonumber
 \end{equation}
 \end{itemize}
 \end{center} 
\end{minipage}

      \begin{center}
      \item -The tunneling probability takes a thermal form:
      \begin{center}
      $\Gamma \propto exp\left(-2Im\mathcal{I}\right) \propto exp(-\omega/T) \Rightarrow T= \omega/2 Im \mathcal{I} \ ,$
      \end{center} 
      \end{center}

\bigskip
\noindent
\begin{minipage}[t]{.5\textwidth}
\raggedright
 \begin{center}
 \begin{itemize}
 \item with temperature:
 \begin{equation}
 T=+\frac{\kappa_{\cal H}}{2\pi} \ . \nonumber
\end{equation} 
 \end{itemize}
 \end{center} 
\end{minipage}
\hfill
\noindent
\begin{minipage}[t]{.5\textwidth}
\raggedleft
 \begin{center}
 \begin{itemize}
 \item with temperature:
 \begin{equation}
 T=-\frac{\kappa_{\cal H}}{2\pi} \ .\nonumber
\end{equation} 
 \end{itemize}
 \end{center} 
\end{minipage}
\bigskip

Therefore, depending on the future/past nature of the horizon, the temperature is proportional to plus/minus the surface gravity. Moreover, we have shown in the above that the Lie derivative $\mathcal{L}_{n.v} \theta_{v}$ (where ``$n.v$'' and ``$v$'' stand for ``direction of non-vanishing/vanishing expansion''), is always opposite in sign to the surface gravity. Thus, outer horizons have positive surface gravity, and inner horizons have negative $\kappa_{\cal H}$. We see that we have two sign effects here, one in the surface gravity, due to the inner/outer nature of the horizon, and one in the temperature, due to the future/past nature of the apparent horizon. Both effects combine to bless the future-outer and past-inner configurations with positive temperature, whereas the other two configurations have negative $T$. We summarize the situation in Table \ref{table_Hawking_temp}.

\bigskip
\begin{table}[h]
\centering
\begin{tabular}{|l|c|c|c|}
  \hline
  $\phantom{rien}$ & inner ($\mathcal{L}_{n.v} \theta_{v}>0$)  & outer ($\mathcal{L}_{n.v} \theta_{v}<0$) & $\phantom{rien}$ \\
  \hline
  future ($\theta_{n.v}<0$) & - & + & $T\propto +\kappa_{\cal H}$ \\
  \hline
  past ($\theta_{n.v}>0$)   & + & - & $T\propto -\kappa_{\cal H}$ \\
  \hline
  $\phantom{rien}$ & $\kappa_{\cal H}<0$ & $\kappa_{\cal H}>0$ &   $\phantom{rien}$ \\
  \hline
\end{tabular}
\caption{Sign of Hawking temperature depending on nature of horizon.}
 \label{table_Hawking_temp}
\end{table}

\section{The four kinds of apparent/trapping horizons, and feasibility of Hawking radiation}
 \label{section_physical_arguments}
 
\begin{figure}[h] 
\centering
  \includegraphics[width=8cm,angle=0]{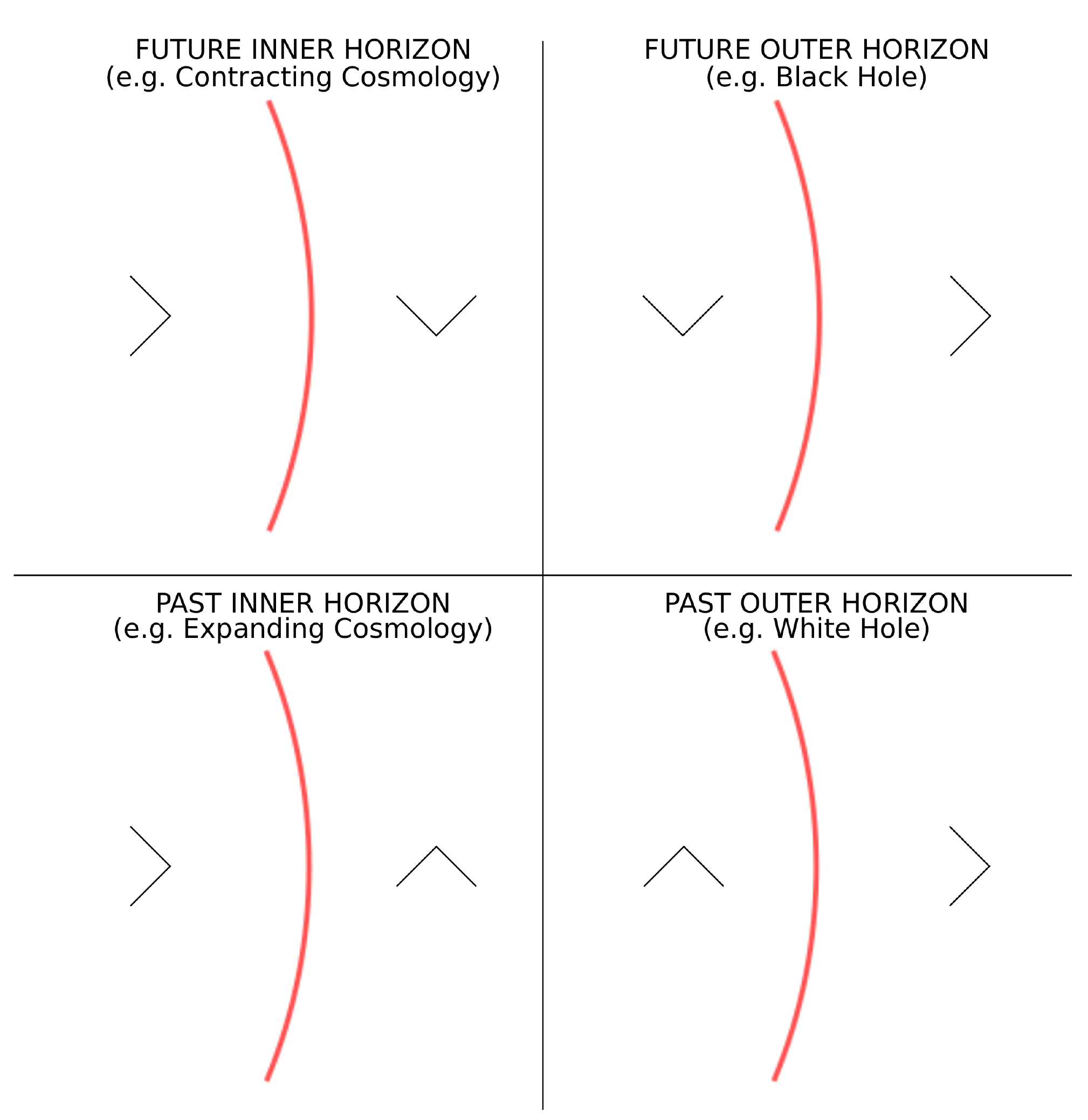}    
   \caption{The four types of apparent/trapping horizons} 
    \label{fig_trapping}
\end{figure}

The previous sections deal with apparent/trapping horizons in a very general way. Let us now apply our conclusions to some precise configurations. Indeed, certain physical systems can be described in the language of trapping horizons.
My claim is that Hawking radiation can occur for other configurations than the future-outer one (\emph{i.e.} the black hole one). To my understanding, the important feature is not the future/past character of the horizon: in both cases one has a well-defined time direction and associated notion of energy. What seems to be important for the existence or not of Hawking radiation is a subtle effect in the combination of the time and spatial characters of the horizon. In the picture of Hawking radiation generated by pair creation at the horizon, this subtle effect must be such that it separates the two members of the pair. In rough terms, if one considers a pair emitted towards the (future) direction where $\theta$ changes its sign, the effect should separate the tunneling particle from the trapped one.

 \subsection{Future-outer trapping horizon: black holes}
\begin{figure}[h] 
\centering
  \includegraphics[width=8cm,angle=0]{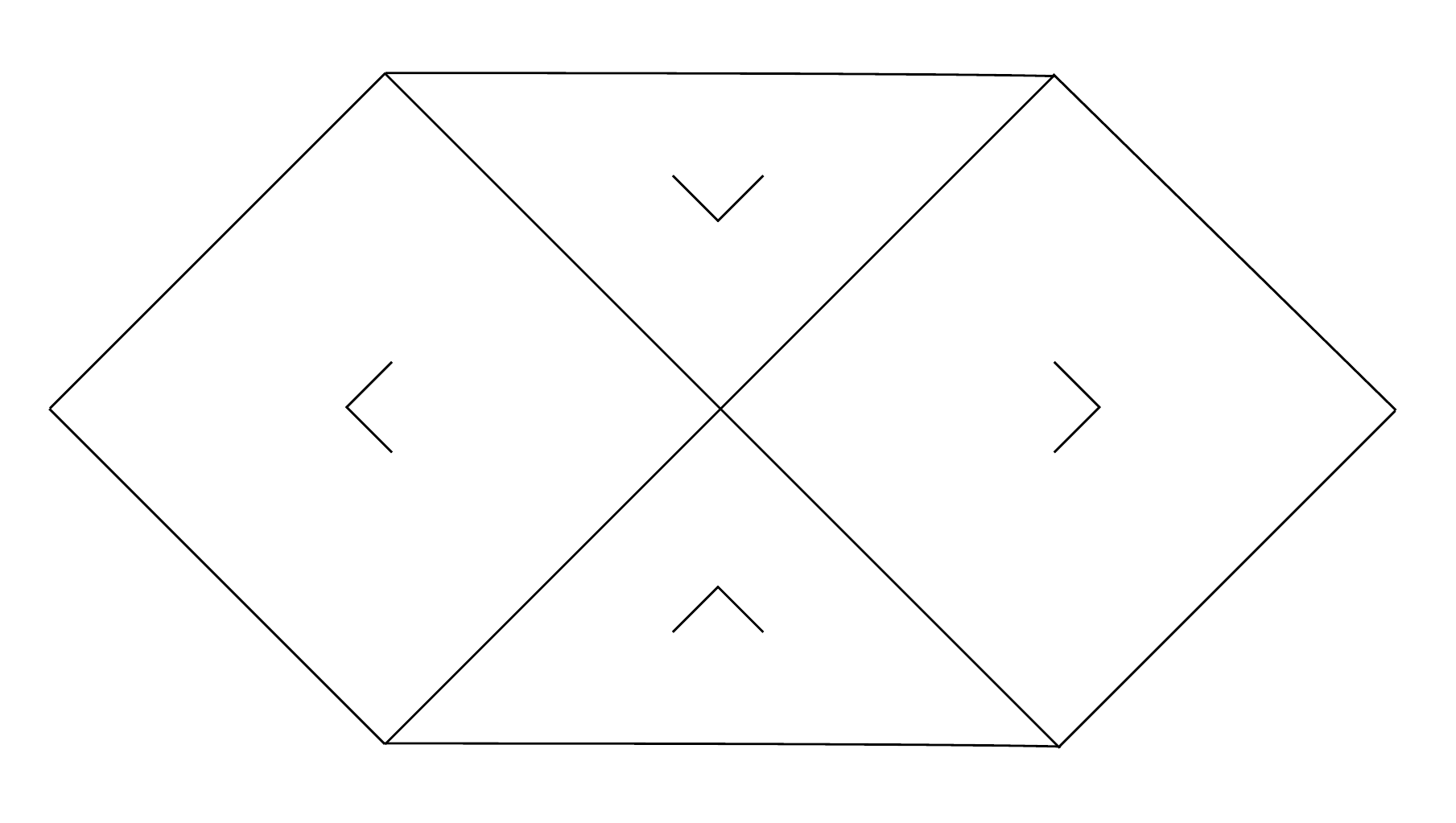}    
   \caption{The maximally extended Schwarzschild solution} 
    \label{fig_Kruskal}
\end{figure}

A black hole can be defined in terms of trapped surfaces. Outside the black hole, an ingoing, spherical light pulse is contracting, whereas an outgoing one is expanding. Inside the black hole region however, both ingoing and outgoing wave-fronts are contracting. Therefore, the future-outer expansion $\theta_+$ is going from positive outside the black hole to negative inside. The apparent/trapping horizon is just the locus where this expansion vanishes. It is such that:
\begin{equation}
\theta_+=0, \quad  \theta_-<0 \quad \text{ and } \quad \mathcal{L}_- \theta_+ < 0 \ ,
\end{equation}
which is the definition of a future-outer trapping horizon. See Figure \ref{fig_Kruskal} for a representation in terms of Bousso wedges. 
Since $\theta_+$ is vanishing, let us consider a pair emitted outwards, at the horizon. One particle is trapped in the black hole region, and cannot travel outwards. The other particle tunnels through the horizon and, being in a normal region of spacetime, can effectively travel outwards, see Figure \ref{fig_trapping_Hawking}. The apparent/trapping horizon acts like a separating membrane. Hawking radiation therefore occurs. 

 \subsection{Past-inner trapping horizon: expanding cosmology}
\begin{figure}[h] 
\centering
  \includegraphics[width=8cm,angle=0]{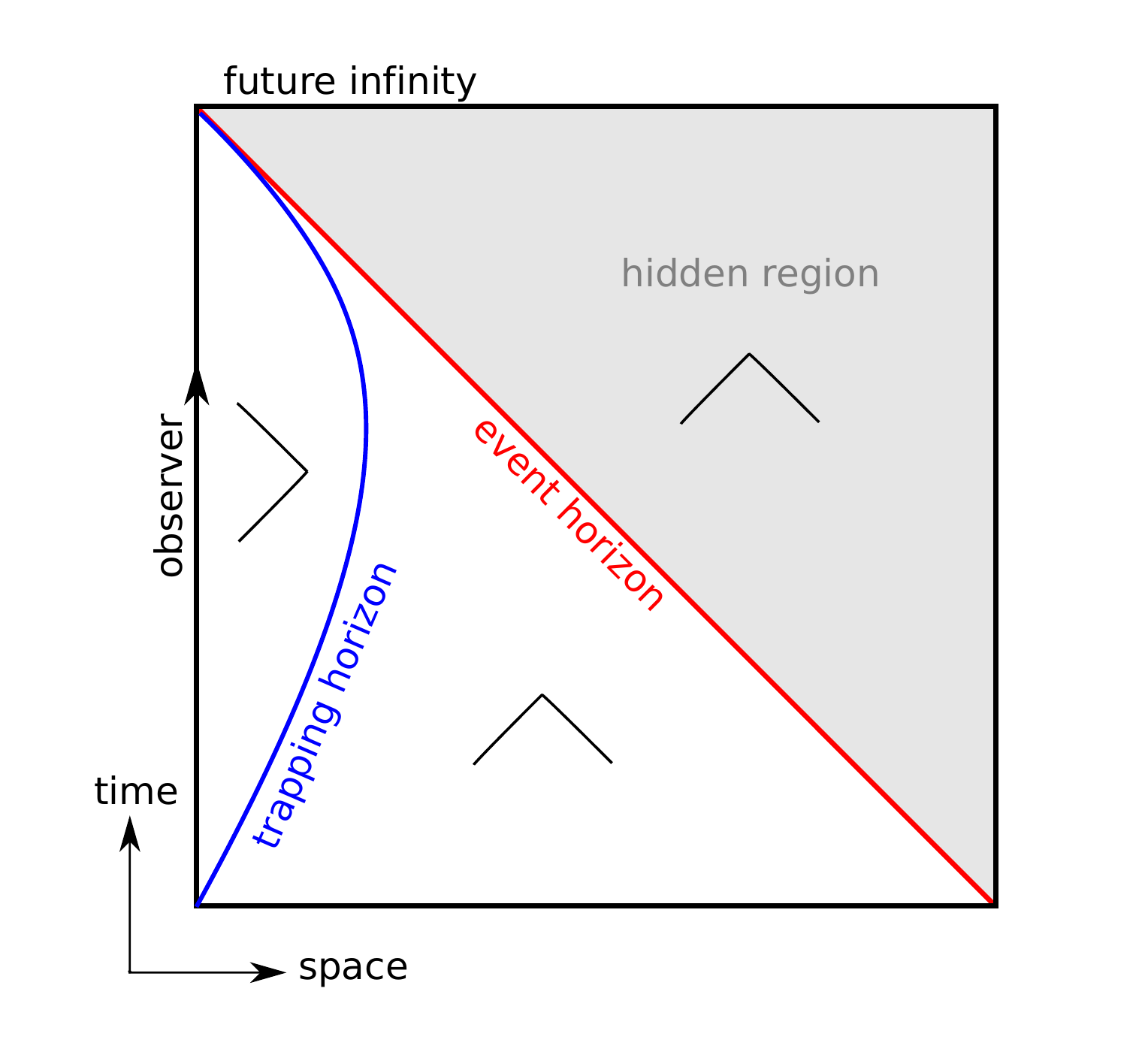}    
   \caption{FLRW Universe with $\Omega_M=0.3$ and $\Omega_{\Lambda}=0.7$} 
    \label{fig_FLRW}
\end{figure}

For the FLRW cosmology, we also have an apparent/trapping horizon, defined as above as the locus where a given expansion $\theta$ is vanishing.
Computing the future inner/outer expansion $\theta$ for the line element of a FLRW Universe: 
\begin{equation}
ds^2 = -d\tau^2 + \frac{a(\tau)^2}{1-kr^2}dr^2 + R^2d\Omega^2 \ ,
\end{equation}
we get:
\begin{equation}
\theta_{-} =  H-\frac{1}{R}\sqrt{1-kr^2} \ ,
\end{equation}
\begin{equation}
\theta_{+} =  H+\frac{1}{R}\sqrt{1-kr^2} \ .
 \label{eq_theta_plus_cosmo}
\end{equation}
In an expanding Universe ($H>0$), the future-inner expansion $\theta_-$ vanishes at some point, which is by definition where the apparent/trapping horizon lies. Therefore in this case, the future-outer expansion has constant sign, but now the future-inner expansion changes sign: from negative in the normal region to positive in the (anti)trapped region. We thus have that: 
\begin{equation}
\theta_-=0, \quad  \theta_+>0  \quad \text{ and } \quad \mathcal{L}_+ \theta_- > 0 \ ,
\end{equation}
which is the definition of a past-inner trapping horizon. In the case of flat space ($k=0$), this trapping horizon is nothing else than the Hubble sphere. See Figure \ref{fig_FLRW}. 
$\theta_-$ vanishes, so we consider a pair emitted inwards at the horizon. The first member of the pair is (anti)trapped in the exterior region and cannot effectively travel inwards, due to a stronger Hubble flow. But its partner may tunnel through the horizon into the normal region, where it will be able to travel inwards. The pair is separated at the horizon. Hawking radiation occurs.

 \subsection{Past-outer trapping horizon: white holes}
Let us now replace our former black hole at the center of the coordinates, by a so-called white-hole. This is also a past-trapped region, as for the expanding cosmology. The difference is that now the normal region is in the outward direction (faraway from the white hole), whereas it was in the inward direction for the FLRW Universe. Outside the white hole, a future outgoing spherical wave-front will expand, and an ingoing one will contract. But inside the white hole region, both future outgoing and ingoing wave-front will expand. Therefore, the future-outer expansion has constant sign, and the future-inner one changes sign, from positive inside the hole to negative outside. This means that, on the horizon:
\begin{equation}
\theta_-=0, \quad  \theta_+>0  \quad \text{ and } \quad \mathcal{L}_+ \theta_- < 0 \ ,
\end{equation}
which is the definition of a past-outer trapping horizon. A white-hole configuration is thus different from an expanding Universe, since it is past-outer and not past-inner. Indeed, the white-hole is the time reversal of the black hole: future-outer becomes past-outer. The expanding cosmology, however, is both the time $\&$ space reversal of the black hole: from future-outer to past-inner. This makes a crucial difference in terms of Hawking radiation. Indeed, since $\theta_-$ is vanishing, let us look at a pair of particles emitted inwards from the horizon. The innermost member will not be able to travel inwards, and will be rejected outwards (due to the white hole region). The outermost particle will try to travel inwards (it is in a normal region). Eventually, the two particles will meet again and annihilate, see Figure \ref{fig_trapping_Hawking}. Therefore, the horizon does not act as a separating boundary in this configuration. Hawking radiation does not occur.

 \subsection{Future-inner trapping horizon: contracting cosmology}
\begin{figure}[h] 
\centering
  \includegraphics[width=8cm,angle=0]{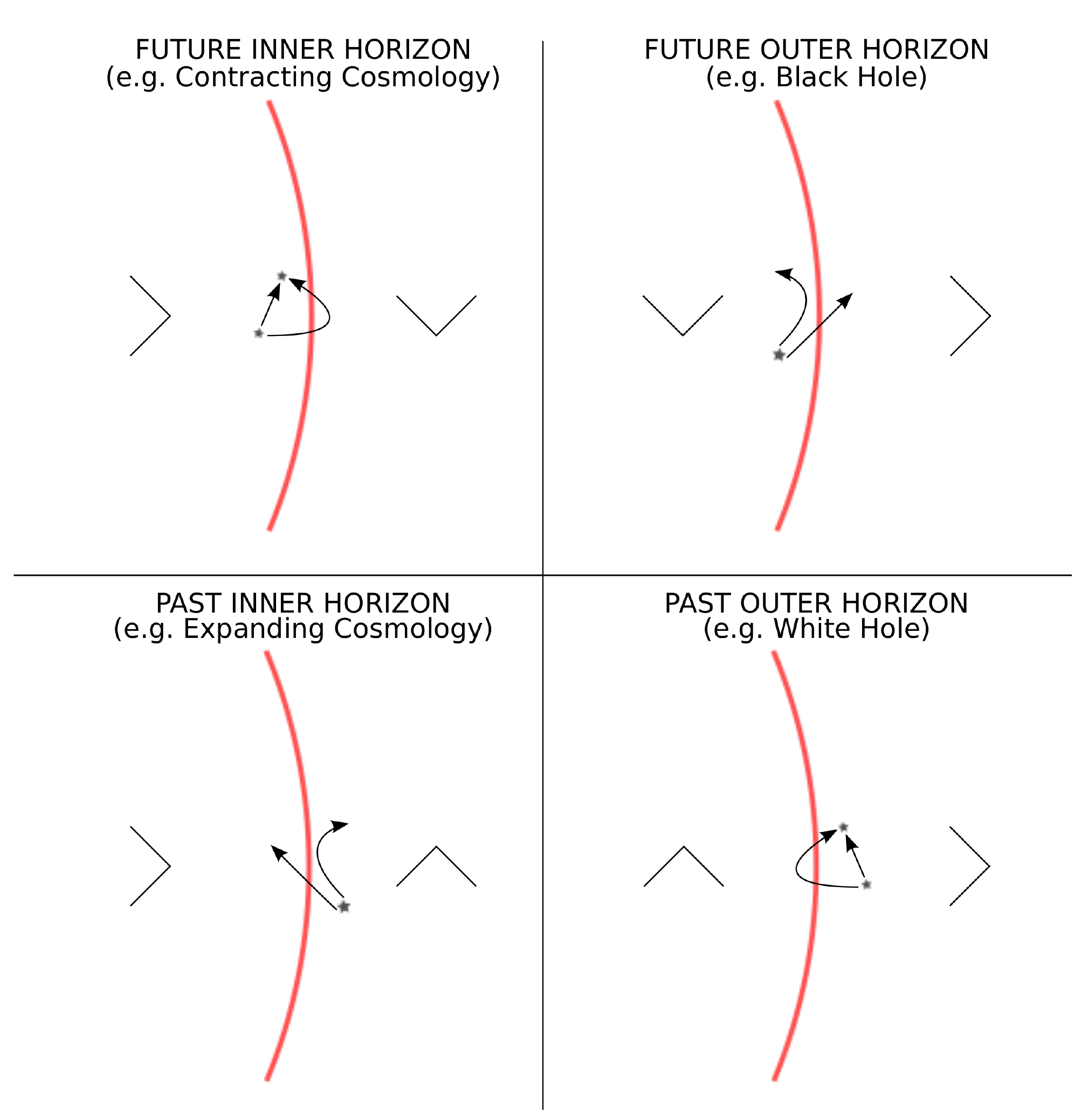}    
   \caption{Feasibility of Hawking radiation for the four types of horizons} 
    \label{fig_trapping_Hawking}
\end{figure}

We consider a contracting cosmology ($H<0$), with an observer at the center of the coordinates. She lies in the normal region, and beyond the horizon lies a future-trapped region. In the normal region, a future ingoing spherical wave-front will contract, and an outgoing one will expand. But beyond the horizon, both future ingoing and outgoing wave-front will contract. It is now the future-outer expansion that changes sign, from negative outside the horizon to positive inside, as seen from Eq.\eqref{eq_theta_plus_cosmo} with negative $H$. This means that, on the horizon:
\begin{equation}
\theta_+=0, \quad  \theta_-<0  \quad \text{ and } \quad \mathcal{L}_- \theta_+ > 0 \ ,
\end{equation}
which is the definition of a future-inner trapping horizon. The vanishing expansion being $\theta_+$, we will focus on a pair of particles emitted outwards. In the normal region (inside the horizon), a pair emitted outwards will travel outwards and annihilate, whereas in the trapped region beyond the horizon, the outward-emitted pair will travel inwards, and annihilate all the same. At the horizon however, the innermost particle of the outward-emitted pair will travel outwards, while the outermost one will travel inwards, due to the trapped region where it is. They will therefore meet and annihilate, see Figure \ref{fig_trapping_Hawking}. There again, the apparent/trapping horizon does not act like a separating membrane: Hawking radiation will not occur. 

Note 1: Since a white-hole is the time-reversal of a black hole, one should expect it to emit Hawking radiation \emph{to the past}. This would mean that the pair is separated towards the past, but is nothing else than saying the particles will meet and annihilate in the future. So I don't think Hawking radiation to the past has any physical effect.

Note 2: the future-inner horizon, for contracting cosmology, may also represent the IMOTS of a black hole. One therefore recovers the conclusions of \cite{Ellis:2013oka}, that the IMOTS does not emit Hawking radiation, whereas the OMOTS does (or at least may).

\section{Conclusion}
 \label{section_conclusion} 
In the present article, we have presented a way of computing well-defined surface gravity and temperature of the horizon, for a wide class of spherically symmetric, dynamical spacetimes. In particular, we stress the fact that signs of these quantities matter: one should not choose one's preferred sign or use absolute values for convenience. 

We have shown that the future/past, inner/outer nature of trapping horizons has a two-fold effect on the sign of their temperature. The first sign effect is that the surface gravity is positive for outer horizons, and negative for inner horizons. Indeed, the expression for $\kappa$ can be linked to the Lie derivative of the relevant expansion $\theta$, the sign of which defines the inner/outer character. 
 The second sign effect is that $T\propto + \kappa$ for future horizons, while $T\propto - \kappa$ for past ones. This comes from the contour integral performed in the tunneling method.
 The final result is that future-outer and past-inner horizons are blessed with a positive temperature, whereas future-inner and past-outer ones have negative $T$. This is understood when considering examples of these four kinds of trapping horizons, and the dynamics of particles in their vicinity.
The behaviour of Hawking pairs in the four different configurations is summarized in Figure \ref{fig_trapping_Hawking}. The physical arguments described in Section \ref{section_physical_arguments} are in full agreement with the computations and summarizing table of Section \ref{section_Hawking_rad_from_tunneling}. We find that, in the two cases where the computed temperature is positive, the horizon effectively acts as a separating boundary for a pair of particles. These are the black hole case and the expanding cosmology. On the other hand, the other two cases yield a negative parameter $T$, and correspond to configurations where the particle/antiparticle pair is not separated by the horizon, but rather forced to recombine and annihilate by the conspiring dynamics of the interior and exterior regions. Hence there should occur no Hawking radiation.

The negative $T$ of the white hole and contracting cosmology cases might not be interpretable as a true temperature, but rather as a breakdown of our computation method (Section \ref{section_Hawking_rad_from_tunneling}). It is possible that the statistical ensemble we work with is not applicable to those two cases, as is often the case when one comes up with ``negative temperature''. Another interpretation would be to see this negative $T$ as a sign of the instability of white holes and contracting cosmologies, maybe in relation with what happens in systems with a population inversion, such as laser cavities.


\vspace{1cm}

\textbf{Acknowledgement:} I thank George Ellis for a discussion which led me to investigate the white hole sector, and eventually gave me the idea of the present paper. I thank Pierre Bin\'{e}truy, Carlos Barcel\'{o}, as well as Gia Dvali and the group at LMU, for useful comments. I thank Ma\"{i}ca Clavel for helping me with Inkscape.

\end{document}